# High performance fully etched isotropic microring resonators in thin-film lithium niobate on insulator platform

**MEISAM BAHADORI,**[1,*] **YANSONG YANG,**[1] **LYNFORD L. GODDARD,**[2] **AND SONGBIN GONG**[1]

[1]*Illinois Integrated RF Microsystems Group, Micro and Nanotechnology Laboratory, Department of Electrical and Computer Engineering, University of Illinois at Urbana-Champaign, Urbana, Illinois 61801, USA*
[2]*Photonic Systems Laboratory, Micro and Nanotechnology Laboratory, Department of Electrical and Computer Engineering, University of Illinois at Urbana-Champaign, Urbana, Illinois 61801, USA*
songbin@illinois.edu
[*]*meisam@illinois.edu*

**Abstract:** We present our design, fabrication, and experimental results for very high-performance isotropic microring resonators with small radii (~ 30 μm) based on single-mode strip waveguides and transverse magnetic (TM) polarization in a fully etched lithium niobate (Z-cut) thin-film on insulator. The loss of the devices is predicted to be < 10 dB/cm, and is measured to be ~ 7 dB/cm. The measured optical responses of microring resonators exhibit an extinction of ~ 25 dB (close to critical coupling), a 3 dB optical bandwidth of 49 pm (~ 6 GHz) for all-pass structures, an extinction of ~ 10 dB for add-drop structures, and a free spectral range of ~ 5.26 nm, all of which are in excellent agreement with the design. This work is the first step towards ultra-compact and fully isotropic optical modulators in thin-film lithium niobate on insulator.

## 1. Introduction

The increasing demand for higher bandwidth in data centers and cloud-based services has inevitably led to the advancement and deployment of optical transceivers [1,2]. Pluggable transceivers based on directly modulated vertical cavity surface emitting lasers (VCSELs) [3,4] are currently dominating the market due to their low cost per bit solution for short-range interconnects (intra data-center). As the size of these communication intensive centers continues to grow, solutions for scalable long-range interconnect are coming into the spotlight [1,5,6].

*Scalability* for photonics requires integration; a lesson well learned from the electronics counterpart. *Long-range* reach requires externally modulated lasers with high enough optical power. Therefore, photonics technologies that can readily take advantage of complementary metal oxide semiconductor (CMOS) foundries have the potential for mass volume production and breaking the cost barrier. Silicon photonics (SiPh) is hailed as the frontier of such technologies [7–9] and has been in development for at least two decades. Thanks to its high index contrast (HIC) property [10], the core photonic functionalities such as optical routing with waveguides [11], low propagation loss (1-3 dB/cm) [12], power splitters and combiners [13–15], wavelength selective filters [16,17], broadband Mach-Zehnder modulators [18,19], wavelength selective modulators [20,21], low-loss vertical grating couplers [22–24] and edge couplers for optical input/output are now well-established and various foundries around the world offer their own standardized process development kits (PDKs) to the public [9].

Although SiPh holds the highest promise in terms of scalability and production cost, its certain drawbacks, such as its reliance on carrier dynamics for modulation as well as the induced excess loss, limit its adoption for niche applications where linear electro-optic modulation or strong nonlinear optical effects are highly desired. Examples include second harmonic generation [25] or transferring equi-level 4-level pulse-amplitude modulation (PAM4) electrical voltages onto optical signal which requires special electrical equalization [26,27] or other techniques to compensate for the nonlinear electro-optic response of silicon phase shifters.

Similar to SiPh, the lithium niobate on insulator (LNOI) platform provides a high index contrast ($\Delta n = 0.7$) for shrinking the size of photonic structures [28]. Moreover, its linear electro-optic response and larger optical transparency window ($\lambda = 0.35-5.2$ µm) equip the LN platform with superior properties to SiPh when very high performance is demanded. Thin-film LNOI (thickness ~ 400–800 nm) is the ideal choice for bringing LN into the photonics integration arena. However, because LN is a challenging material to etch, hybrid (heterogeneous) solutions [29] were initially adopted. Several types of hybrid waveguides and modulators have been proposed and demonstrated with silicon [30–32], silicon nitride [33], and tantalum pentoxide [34] materials. Because the optical indices of these materials are close to or higher than LN, most of the optical confinement occurs outside of LN which poses a significant detrimental effect to the electro-optic performance of such modulators. To overcome this and induce more optical confinement in LN, partially etched LN has been recently proposed to create optical waveguides as a path to realizing high-performance electro-optic traveling wave structures [35] and resonant devices [36–38]. Due to the optical anisotropy of LN [39], the performance of the demonstrated structures based on X-cut or Y-cut crystals (in-plane extraordinary axis) [40] and TE optical mode (in-plane polarization) has an inherent dependence on their orientation. This is especially pronounced for microring-based structures [37], where their optical or electro-optic response can significantly degrade if the structures have some in-plane rotation or misplacement. This problem also exists for vertical grating couplers that have been so far demonstrated in LN platforms [41,42]. Although hybrid Si-LN grating couplers [43] might be a solution to this problem, their fabrication is more complex and challenging. Vertical grating couplers provide several advantages including compactness and the ability to couple light from anywhere on the surface of the chip (e.g., flip-chipped VCSELs), thereby enabling rapid die/wafer testing and heterogeneous integration with light sources. Natually, the aforementioned restrictions on the orientation of grating couplers are undesirable.

In this work, we demonstrate isotropic grating couplers and isotropic low-loss microring structures by utilizing TM polarization in Z-cut LN crystal. The optical extraordinary axis is normal to the plane of the chip; hence the in-plane axes are both ordinary (with the same optical index) and no sensitivity to the device orientation occurs. Furthermore, because TM polarization has less overlap with the sidewalls than its TE counterpart, our fabricated fully etched rectangular single mode waveguides (~ 800 nm wide and 560 nm thick) exhibit low propagation loss (~ 7 dB/cm for circular waveguides). By carefully designing the optical mode and the coupling between microrings and straight waveguides we demonstrate very high resonance extinction (~ 25 dB) for a very small radius (~ 30 µm) and high $Q$-factor (~ 31000). The extinction for our thin-film LN microring resonators is higher than previous results in the literature, e.g., ~ 5−8 dB was reported in [36], ~ 12 dB was reported in [42], ~ 4 dB was reported in [40], and ~ 6.5 dB was reported in [37]. Our work will pave the way for realizing ultra-compact and isotropic microring modulators with speeds of ~ 10 Gb/s in thin-film LNOI.

## 2. Fabrication process

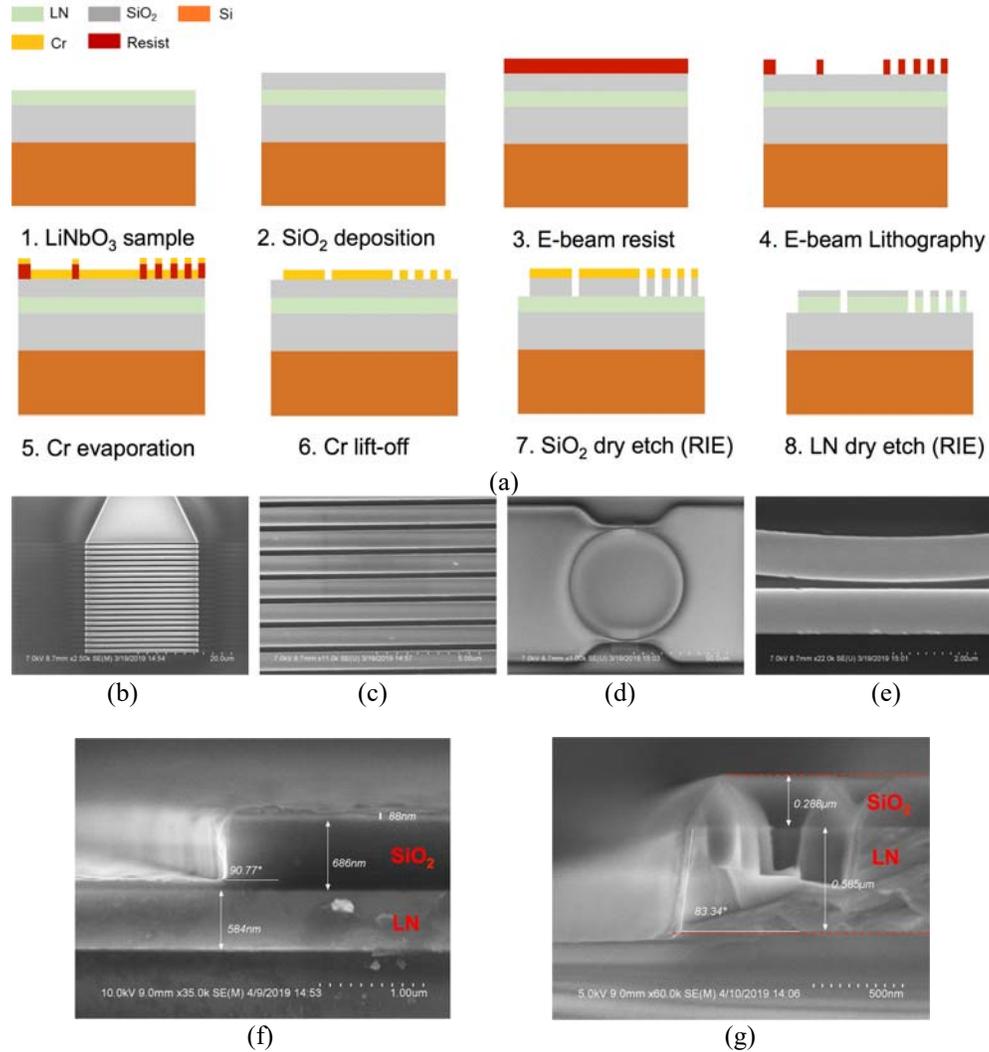

Fig. 1. (a) Fabricaton steps. SiO$_2$ is used as a hard mask for etching LN. Cr is chosen as a hard mask for defining SiO$_2$. (b)-(e) SEM images of various devices after the Cr deposition (step 6). (f) Side view SEM image of etching the top SiO$_2$ in a dummy sample (g) Sideview SEM image of a fully etched LN thin-film, showing a sidewall angle of 83°.

We fabricated devices using the process steps shown in Fig. 1 (a). The Z-cut thin-film lithium niobate wafers (nominal thickness of 560 nm) on 2.5 μm SiO$_2$ on a silicon substrate were purchased from a commercial vendor (NGK Insulators, LTD.). The thickness of the buried oxide was chosen such that the leakage of optical mode into the substrate for both TE and TM polarizations is negligible. A ~ 700 nm layer of SiO$_2$ is chosen as the hard mask for dry-etching LN and deposited using standard PECVD (~ 113 nm/min). 300 nm of photoresist (PMMA) was spin coated at 2000 rpm for 1 minute and electron beam lithography (EBL) was performed on the sample. Proximity effect correction was performed using Beamer software to generate a dose table that enabled faithful patterning of the smallest feature sizes of our designs (300–400 nm). A 120-nm thick chromium (Cr) layer patterned via evaporation and lift-off process was chosen as the mask for etching $SiO_2$ ($SiO_2$:Cr rate ~ 20:1). Figures 1 (b)-(e) show the scanning electron microscope (SEM) images of the patterned Cr for various devices such as grating couplers [Fig. 1(b) and (c)] and add-drop microring resonators [Fig. 1(d) and (e)]. Next, the $SiO_2$ layer and then the LN layer were dry etched. The dry etching of $SiO_2$ and LN is done using Plasma-Therm ICP-RIE system. The $SiO_2$ was etched using $CHF_3$ (20 sccm) and $O_2$ (5 sccm) gases with ICP power of 800 W and RIE power of 100 W. The LN was etched using $Cl_2$ (5 sccm), $BCl_3$ (15 sccm), and Ar (18 sccm) gases with ICP and RIE powers of 800 W and 280 W, respectively. The etch rates of $SiO_2$ and LN in the LN etching are 140 nm/min and 200 nm/min respectively, corresponding to the selectivity of 1.45:1. The thickness of the remaining $SiO_2$ after fully etching LN was measured to be ~ 280 nm. This thickness was left on top of LN layer intentionally to improve the coupling loss of grating couplers [41]. Figure 1(f) shows the side view SEM image of the etched $SiO_2$ and Fig. 1(g) shows the side view SEM image of etched LN thin-film. The angle of LN sidewalls is estimated to be 83°.

## 3. Design of microring resonators

In this section, we provide details of designing our microring resonators, including the choice of an optical waveguide for single-mode operation, choice of radius for the largest free spectral range (FSR), and the design of coupling between the ring and waveguides.

### 3.1 *Optical mode*

Unlike other demonstrated LNOI waveguides [37] that have a large cross-section and can potentially support multi-mode operation, our goal is to design single-mode waveguides. A thickness of 560 nm was chosen based on our optimization of the grating couplers, and the maximal width for single mode operation of the fully etched waveguides was determined by sweeping the width. For a thickness of 560 nm, the maximal width for single-mode operation was determined to be 1000 nm; however, we selected a design with an 800 nm width to achieve stronger coupling. Because isotropic and low-loss operation is targeted, TM polarization was chosen to minimize the overlap of the optical mode with the sidewalls, as well as enabling maximum electro-optic efficiency operation in the Z-cut crystal for our future designs. Figure 2(a) shows the calculated TM effective index of the 800 nm × 560 nm waveguide as a function of the wavelength and Fig. 2(b) shows the group index of the mode. At $\lambda$ = 1550 nm, the effective index is 1.736 and the group index is 2.3614. Our previous measurement of Z-cut ring resonators for TE polarization revealed the waveguide loss to be ~ 11 dB/cm [44,45]. Therefore, a lower loss for TM polarization is expected (~ 5–10 dB/cm).

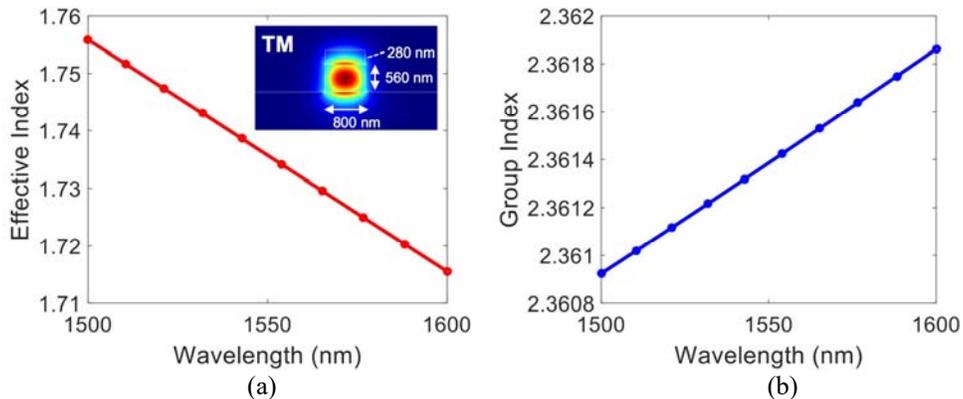

Fig. 2. (a) Calculated effective index of the fundamental TM mode in a fully etched waveguide of 800 nm × 560 nm in dimensions. (b) Calculated group index of the fundamental TM mode.

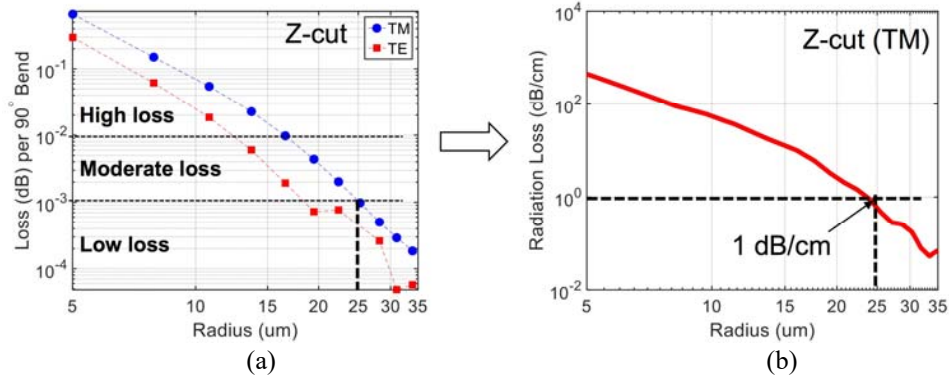

Fig. 3. (a) Calculated bending loss (radiation + mode-mismatch) for a 90° bend for TE and TM polarizations. (b) Estimated radiation loss (dB/cm) as a function of the radius. Losses < 1 dB/cm can be achieved for a radius > 25 μm.

### 3.2 Bending loss and choice of radius

In order to achieve the smallest microring resonator and hence the largest FSR, we calculated the bending loss of our waveguides. Figure 3(a) shows the estimated bending loss (radiation loss plus the mode-mismatch loss [46–48]) per 90° bend for both the TE and TM polarizations in our waveguide. As expected, the TM polarization shows a slightly higher loss due to its lower confinement. Because the mode-mismatch loss is not present in a fully circular ring, we extract only the radiation loss by comparing the loss of a 90° bend to the loss of a 180° bend. The result (in dB/cm unit) is plotted in Fig. 3(b) for the TM polarization. We see that 1 dB/cm loss can be achieved at a radius of 25 μm. Hence, we choose the radius to be 30 μm which results in a negligible radiation loss (~ 0.2 dB/cm).

Based on the group index and the radius, the FSR is estimated from

$$FSR_{nm} = \frac{\lambda_{nm}^2}{2\pi R_{nm}\, n_g} \tag{1}$$

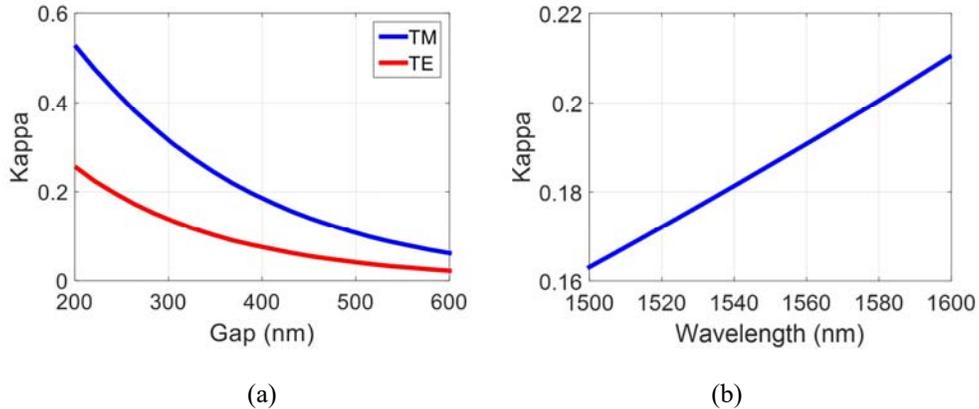

Fig. 4. (a) FDTD simulation results of the coupling of Z-cut waveguide to a microring resonator with a 30 μm radius at $\lambda$ = 1550 nm for both TE and TM polarizations. (b) Calculated coupling for TM polarization as function of wavelength for gap = 400 nm.

to be ~ 5.26 nm around $\lambda$ = 1550 nm. Such a large FSR has only been recently reported by Krasnokutska *et al.* in thin-film LNOI [36] albeit with rib waveguides of 500 nm in thickness.

### 3.3 Coupling

Based on the predicted loss of < 10 dB/cm for the TM ring resonators, we designed the coupling between the Z-cut waveguide and ring by running an FDTD simulation and sweeping the coupling gap. An analytic approach based on optical supermodes [16] was also utilized to estimate the coupling coefficients. Figure 4(a) shows the simulation results of the field cross coupling (kappa) as a function of the coupling gap at $\lambda$ = 1550 nm. Figure 4(b) shows the cross coupling for TM polarization as a function of the wavelength for a 400 nm coupling gap (measured gap in the fabricated structures). At $\lambda$ = 1550 nm, the field coupling is estimated to be ~ 0.19 and the power coupling is 3.6%.

### 3.4 All-Pass ring resonators

The spectra of the all-pass ring structures are estimated from
$$T(\lambda) = \frac{t^2 + A - 2t\sqrt{A}\cos(\phi)}{1 + t^2 A - 2t\sqrt{A}\cos(\phi)} \tag{2}$$
where $t = \sqrt{1 - \kappa^2(\lambda)}$, $A$ is the round-trip power attenuation, and $\phi = \frac{2\pi}{\lambda} n_{eff}(\lambda) \times 2\pi R$ is the round-trip phase accumulation inside the ring. The critical coupling condition occurs close to resonances for which $t \approx \sqrt{A}$. Figure 5(a) shows the calculated spectrum by assuming 5 dB/cm loss inside the ring and a coupling gap of 400 nm. Figure 5(b) shows the calculated spectrum with 10 dB/cm loss. The non-equal resonance extinctions are due to the dependence of the coupling coefficient on wavelength. For 5 dB/cm loss, the critical coupling happens close to $\lambda$ = 1550 nm, and for 10 dB/cm loss, the critical coupling takes place beyond $\lambda$ = 1600 nm. Based on our predicted loss of < 10 dB/cm, we anticipate that this choice of the gap size (400 nm) will result in the critical coupling condition occuring in the 1500 – 1600 nm wavelength range.

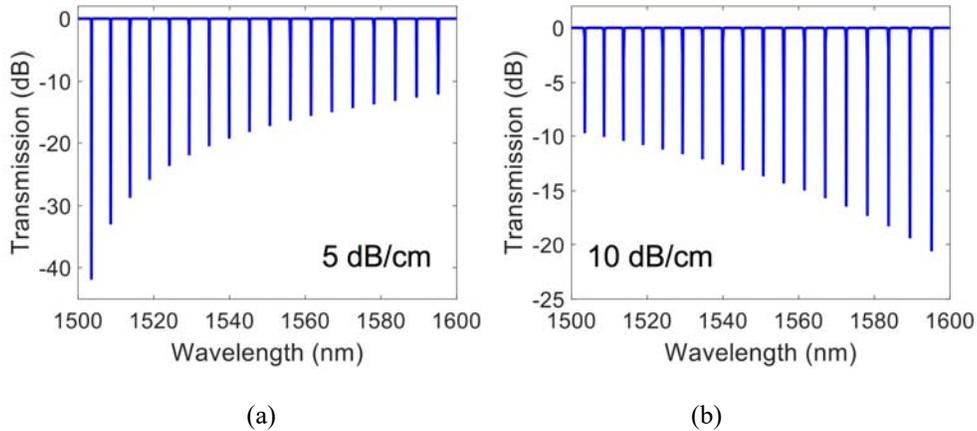

Fig. 5. (a) Calculated spectrum of the all-pass microring structure with 5 dB/cm loss and a 400 nm coupling gap. (b) Calculated spectrum with 10 dB/cm loss.

### 3.5 Add-drop ring resonators

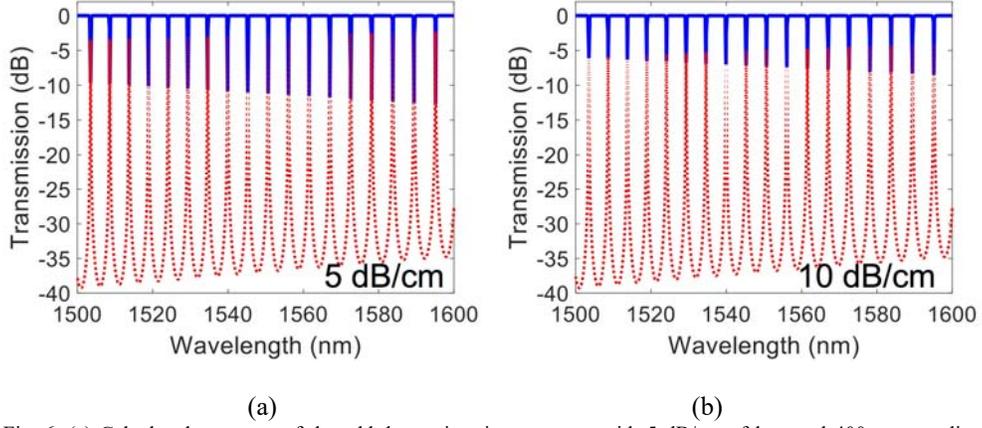

(a) (b)

Fig. 6. (a) Calculated spectrum of the add-drop microring structure with 5 dB/cm of loss and 400-nm coupling gaps. (b) Calculated spectrum with 10 dB/cm of loss.

The add-drop ring resonators are designed with equal gaps at the input and drop ports so that the structures are perfectly symmetric. This ensures that the round-trip loss and the coupling coefficients can be extracted from the drop spectrum without any ambiguity [16,49]. The spectra of add-drop structures are calculated from

$$T(\lambda) = \frac{\kappa_1^2 \kappa_2^2 \sqrt{A}}{1+t_1^2 t_2^2 A - 2t_1 t_2 \sqrt{A} \cos(\phi)} \quad (3)$$

[50] where $\kappa_1$ and $\kappa_2$ are the coupling coefficients of the input and drop ports, respectively.

Figure 6(a) shows the simulated spectra of the add-drop ring with 400-nm gaps and 5 dB/cm of loss. The extinction of the through path is ~ 10 dB and the extinction of drop path is ~ 37 dB. Figure 6(b) shows the simulated spectra for 10 dB/cm of loss, with which the extinction of the through path is reduced to ~ 5 dB and the extinction of the drop path is ~ 33 dB.

## 4. Results and discussions

In this section, we present the fabricated structures and the measurement results as well as analysis of the acquired data to confirm the loss and coupling coefficients of the ring resonators.

Figure 7(a) shows a close-up image of the measuring stage with our sample (1 inch × 1 inch)

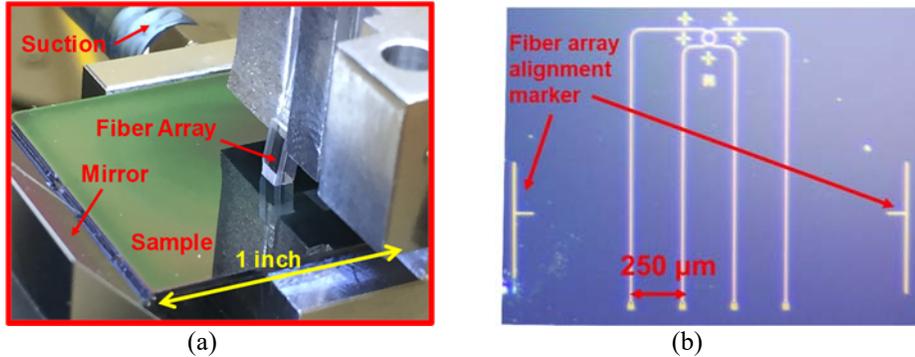

(a) (b)

Fig. 7. (a) Image of the measuring stage. The 1" × 1" sample is mounted on an XYZ stage and a fiber array (four fibers with a 250-μm pitch) is used. Side mirrors are used to view the height of the fiber array during alignment. (b) Optical microscope image of an add-drop structure on the chip. The fiber alignment markers are also shown.

mounted on an XYZ stage with vacuum suction. The fabricated sample has several all-pass structures and several add-drop structures with various coupling gap sizes. A four-fiber V-groove array with a pitch of 250 μm is used for coupling light from single-mode fibers into the grating couplers. The vertical alignment is assisted with the side mirrors, and the in-plane alignment is assisted with the alignment markers that are placed on the LN sample. The marker positions are based on the dimensions of the fiber array; see Fig. 7(b). A continuous wave (CW) tunable laser (Santec 710 TSL) accompanied by a polarization controller (PC) is used to control the wavelength, power, and polarization of the light inside the incoming fiber. The output lights from the through path and the drop path go to photodiodes. The generated photocurrents are amplified by log-amplifier circuits and voltages are read out by a National Instrument data acquisition device (NI DAQ) and imported into LabView software. The output power of the laser is set to 10 dBm, and the sweep range is set from 1500 nm to 1600 nm. The number of samples is 50,000 which yields a resolution of 2 pm for the sweep. This sweep resolution is sufficient to easily resolve resonances with quality factors less than 50,000. The polarization controller is set such that the output power from the through path of the structures is maximized at an off-resonance wavelength.

### 4.1 Compact grating couplers for optical I/O

Vertical grating couplers were designed for TM polarization and fabricated to enable optical coupling to the chip. Figure 8(a) shows the SEM image of an array of four grating couplers with a zoomed-in view of the first grating. The pitch of the grating is 1.127 μm after fabrication, and the duty cycle is 50.4%. Figure 8(b) shows the coupling coefficient of the grating coupler extracted from the transmission of two back-to-back gratings. Each grating exhibits ~ 16.8 dB of loss at 1550 nm which is worse than the design due to fabrication errors that alterned the pitch and duty cycle from their design values. The 1-dB bandwidth of the fabricated gratings is > 60 nm, and the 3-dB bandwidth is > 100 nm. High coupling loss is expected from our grating couplers because they are all fully etched. Grating couplers with a partial etch exhibit better coupling coefficients [41].

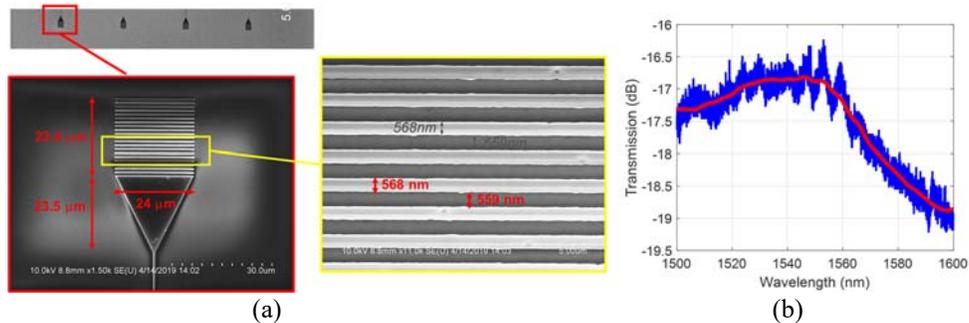

(a)            (b)
Fig. 8. (a) Top view of an array of four grating couplers and zoomed-in view of the first grating. The pitch of fabricated grating is 1.127 μm and the number of grating periods is 20. (b) Measured coupling coefficient of one grating coupler and the smoothed data (red curve) as a guide for the eye.

### 4.2 All-Pass microring resonators

The SEM image of a fabricated all-pass structure is shown in Fig. 9(a). Two back-to-back couplers are also included to extract the coupling loss of grating couplers. Figure 9(b) shows the zoomed-in view of the ring and waveguide. Figure 9(c) shows that the coupling gap is ~ 420 nm which is 20 nm larger than the design value (400 nm). Figure 9(d) shows the layout of the structure indicating that the light was coupled into the chip from port 1 and was measured at port 4. The measured spectrum is plotted in Fig. 9(e). It is seen that the resonances of this devices show very high extinction (15–25 dB) with the highest extinction (~ 25 dB) around 1530 nm. Figure 9(f) shows a denser sweep with a sweep resolution of 0.2 pm on the two resonances with the highest extinction. The measured FSR is ~ 5.225 nm, which is slightly smaller than the design value of 5.26 nm. This reduction indicates that the width of the LN waveguides might be slightly smaller than 800 nm (estimated to be 790 nm).

Next, we proceed to extract the spectral parameters (3-dB bandwidth, $Q$ factor, kappa, round trip loss) by matching Eq. (2) to the resonances at 1530 nm and 1535 nm, which have the highest extinction. Figure 10(a) shows the result for the resonance close to 1530 nm. The extracted 3-dB bandwidth is 0.0492 nm (~ 6 GHz) and the quality factor is 31,000. The coupling coefficient is 0.166, which is very close to the analytic prediction of 0.165 at 1530 nm for TM polarization. The extracted loss of the ring is ~ 7 dB/cm, which can be considered a reliable estimate of the loss due to the excellent match of the coupling coefficient. If we assume that this resonance is exactly at the critical coupling, the condition $t = \sqrt{A}$ results in a loss of 6.4 dB/cm. This result is in good agreement with the transmission spectra plotted in Fig. 5.

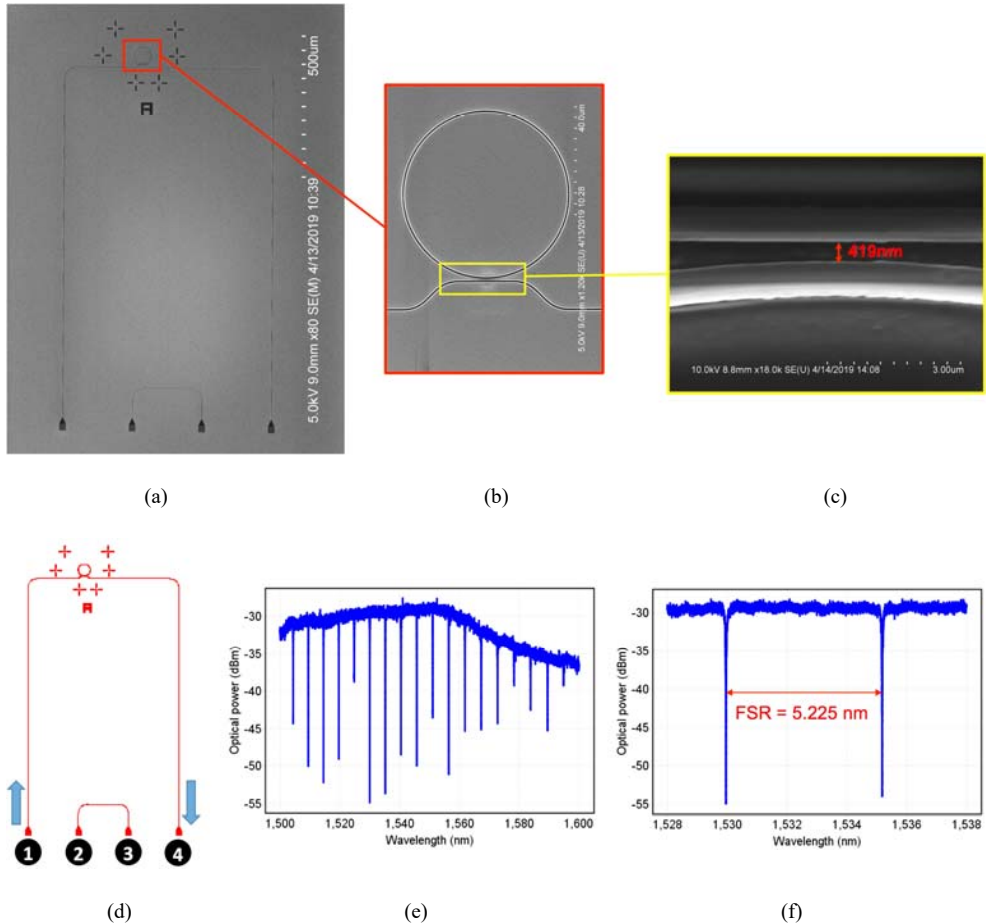

Fig. 9. (a) SEM image of the fabricated all-pass ring resonator. Two back-to-back grating couplers are also included to extract the coupling loss of gratings. (b) SEM image of the ring resonator. (c) SEM image of the coupling gap between the ring and the waveguide. The coupling gap is estimated to be 420 nm. (d) Layout of the structure with the input and output paths of light. (e) Measured spectrum indicating 25 dB of extinction. (f) Zoomed-in view of the two resonances with the highest extinction. The FSR is measured to be 5.225 nm.

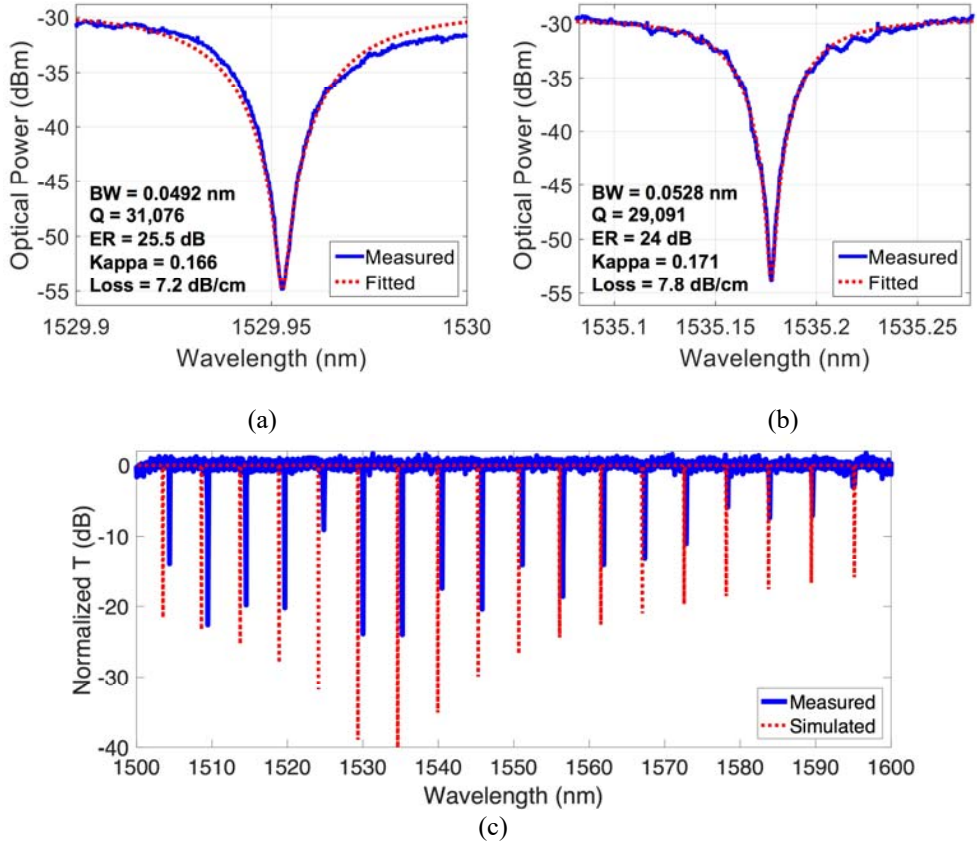

Fig. 10. Extraction of the resonance attributes close to (a) 1530 nm and (b) 1535 nm. Both resonances exhibit high extinction (~ 25 dB). (c) Comparison of the normalized measured spectrum with the simulated one, assuming that critical coupling occurs around $\lambda = 1530$ nm (loss ~ 6.5 dB/cm).

Figure 10(b) shows the result of Lorentzian fitting to the resonance close to 1535 nm. The extracted bandwidth has slightly increased to 0.0528 nm (~ 6.6 GHz), and the coupling coefficient has increased to 0.171, which is in agreement with the analytic value of 0.168. As expected, the coupling coefficient increases with the wavelength. The extracted loss is ~ 7.8 dB/cm which is slightly higher than the previous resonance. We take an average value of 7.5 dB/cm as the loss of the fabricated microring resonator.

Finally, we compare the normalized measured spectrum (by removing the envelope due to the optical responses of the grating couplers) with the simulated spectrum assuming that the critical coupling occurs close to $\lambda = 1530$ nm. The results are plotted in Fig. 10(c). Although the measured spectrum shows smaller extinction than the simulated one, the reduction in the extinction of resonances away from $\lambda = 1530$ nm on both sides agrees well with the observed behavior in the measured spectrum. Therefore, we conclude that our fully etched LNOI microring device has a good performance and operates in accordance with the design targets.

*4.3 Add-Drop microring resonator*

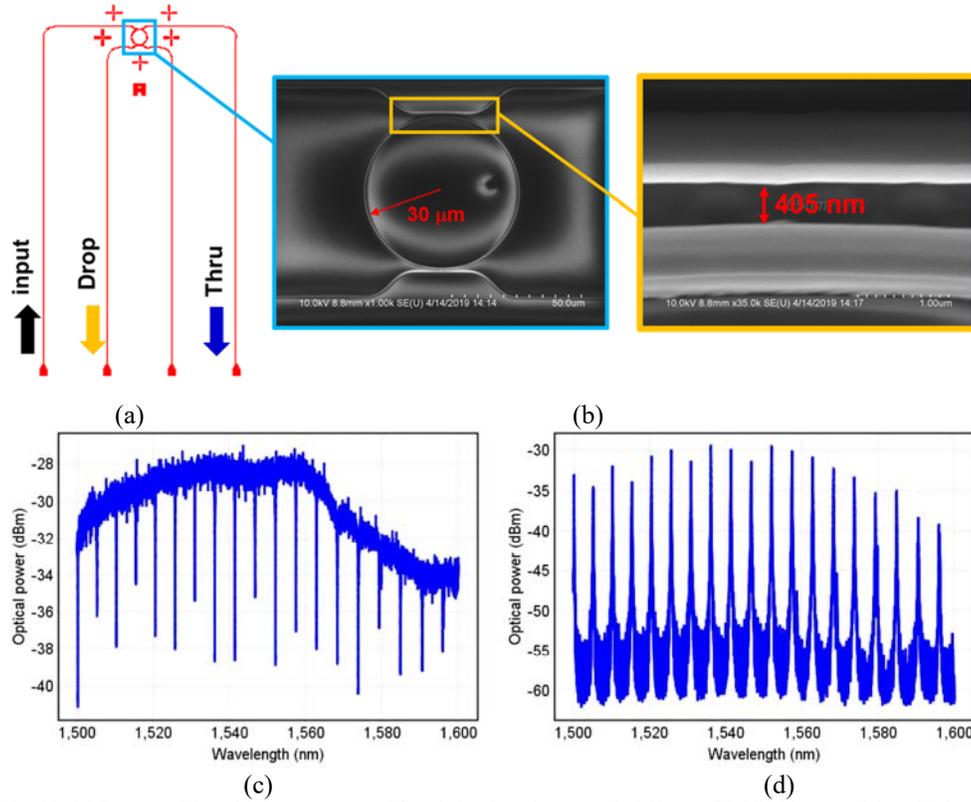

Fig. 11. (a) Layout of the add-drop structure. The pitch of grating array is 250 µm. (b) SEM image of the add-drop microring resonator with a zoomed-in view of the coupling gap between the top waveguide and the ring. The size of the gap is estimated to be 405 nm. (c) Measured spectrum of the through path. (d) Measured spectrum of the drop path.

Symmetric add-drop structures were also designed for TM polarization and fabricated to further investigate the impact of coupling on the spectral response. Because symmetric add-drop structures never reach the critical coupling condition, extraction of coupling coefficients and the round-trip loss of the ring from the drop spectrum is possible. Figure 11(a) shows the layout of the add-drop structure with 400-nm coupling gaps at the input and drop ports. The fiber array is aligned with the four gratings and the drop and through spectra are simultaneously captured by sweeping the tunable laser with 2-pm resolution. Figure 11(b) shows the SEM image of the add-drop microring resonators with a zoomed-in image of the coupling gap between the waveguide and the ring. The coupling gap is estimated to be 405 nm, which is very close to the original design of 400 nm. Figure 11(c) shows the captured spectrum of the through path and Fig. 11(d) shows the spectrum of the drop path. The through spectrum exhibits an extinction of ~ 10 dB while the drop spectrum has an extinction of ~ 30 dB. An extinction much less than 25 dB was expected for the through path because the all-pass structure was close to the critical coupling whereas the add-drop structures include an extra coupling that sets it farther from the critical coupling operation. The measured FSR of the add-drop structure is 5.268 nm, which agrees well with the analytical estimation of 5.26 nm for the 800 nm × 560 nm waveguide.

We next proceed with matching the Lorentzian response of Eq. (3) to the measured spectrum assuming that the input coupling and drop coupling coefficients are identical. The result of the matching is shown in Figs. 12(a) and 12(b) where a good agreement is observed. The bandwidth of the drop resonance was extracted to be ~ 0.108 nm (~ 13.5 GHz) which is approximately twice the bandwidth of its all-pass counterpart (the measured ratio is ~ 2.16). This is expected because the bandwidth of the all-pass structure is given by

$$FWHM_{all-pass} \approx \frac{FSR_{nm}}{2\pi}(2\pi R\,\alpha + \kappa^2) \qquad (4)$$

and the bandwidth of the add-drop structure is given by

$$FWHM_{add-drop} \approx \frac{FSR_{nm}}{2\pi}(2\pi R\,\alpha + \kappa_1^2 + \kappa_2^2) \qquad (5)$$

where $\alpha$ is the power attenuation inside the ring (units of cm$^{-1}$). Therefore, setting $\alpha \approx 1.65$ cm$^{-1}$ (equivalent to 7 dB/cm), $\kappa = 0.166$ (simulated coupling of the all-pass ring with a 420-nm gap), $\kappa_1 = \kappa_2 \approx 0.2$ (simulated coupling of add-drop with a 400-nm gap), and the respective measured FSR values (5.225 nm for the all-pass and 5.268 nm for the add-drop) yields a bandwidth ratio of 2.05 which again agrees well with the measured ratio.

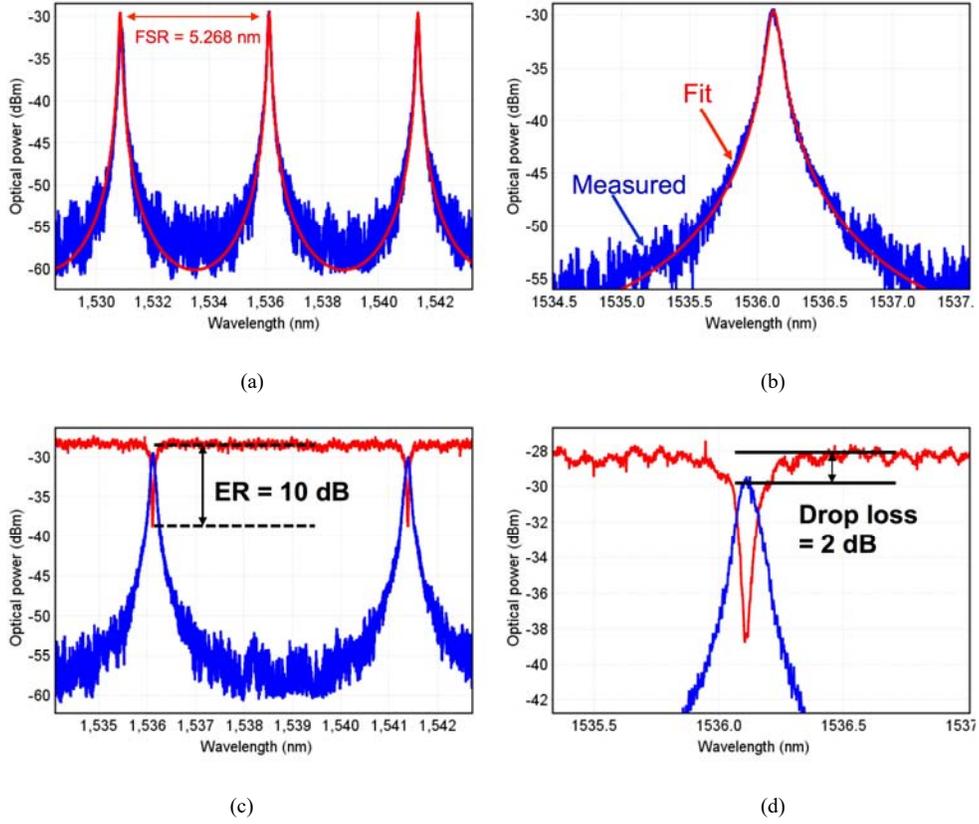

Fig. 12. (a) Measured drop spectrum (blue) and the fitted curve (red). The FSR is ~5.268 nm. (b) Zoomed-in view of the resonances of the drop spectrum. (c) Plot of both the through spectrum and drop spectrum. The extinction of the through path is ~ 10 dB. (d) Zoomed-in view of the through and drop spectra. The relative loss of the drop path at the resonance is ~ 2 dB.

Figure 12(c) shows the spectrum of both through and drop paths indicating ~ 10 dB of extinction for the through path. This also agrees well with the plot presented in Fig. 6(a). Figure 12(d) shows that the relative loss of drop path compared to the through path is ~ 2 dB which is slightly better than the predicted 3 dB of loss in the simulations.

Finally, we remove the grating response envelope in the measured spectrum of the through path in Fig. 11(c) and compare it to the simulated spectrum with ~ 7 dB/cm of round-trip loss. The result is plotted in Fig. 13, showing that the observed ~ 10 dB of extinction is expected from the simulations.

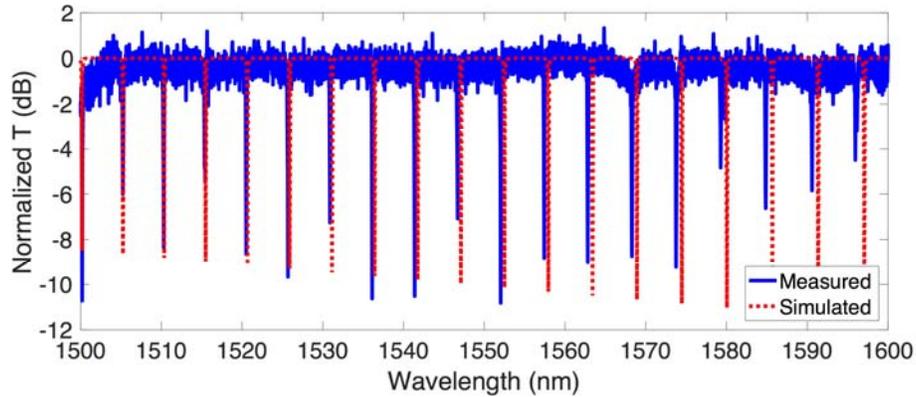

Fig. 13. Comparison of the normalized measured spectrum of the through path of the add-drop microring resonator with the simulated spectrum. A good agreement is observed in the extinctions of the resonances (~ 10 dB).

## 5. Conclusions

We designed fully isotropic (no sensitivity to in-plane orientation) microring resonators in thin-film lithium niobate (Z-cut) on insulator in the form of all-pass and add-drop structures based on *fully-etched single-mode* (TM polarization) strip waveguides (800 nm × 560 nm). Vertical grating couplers were also designed and fabricated on the chip to enable optical input/output. The observed loss of microring resonators was ~ 7 dB/cm with the critical coupling occurring around a coupling gap of 400 nm between the ring and waveguide. Our fabricated all-pass structures demonstrated an extinction of 25 dB (close to critical coupling), optical bandwidth of ~ 6 GHz with an FSR of ~ 5.23 nm while the add-drop structures had an extinction of ~ 10 dB and bandwidth of ~ 13 GHz. The predicted spectral parameters of the microring resonators match well with the measured results indicating the consistency in our fabrication and the high accuracy of our modeling. This work establishes the feasibility of compact and fully isotropic optical modulators in thin-film lithium niobate.

## 6. Funding, acknowledgments, and disclosures

*Funding*

NASA Early Career Award (ECF) (contract number 80NSSC17K052)

*Acknowledgments*

The authors thank Edmond Chow from Micro and Nanotechnology Laboratory (MNTL) at the University of Illinois for his help with the electron beam lithography.

*Disclosures*

The authors declare that there are no conflicts of interest related to this article.

**References**

1. X. Song, R. Li, G. Mi, J. Suo, Z. Zhang, and Y. Li, "Optoelectronic integrated circuits for growing datacenters: challenge, strategy, and evolution," in *Smart Photonic and Optoelectronic Integrated Circuits XXI*, E.-H. Lee and S. He, eds. (SPIE, 2019), p. 11.
2. Q. Cheng, M. Bahadori, M. Glick, S. Rumley, and K. Bergman, "Recent advances in optical technologies for data centers: a review," Optica **5**(11), 1354–1370 (2018).
3. D. Kuchta, A. Rylyakov, C. Schow, J. Proesel, F. Doany, C. W. Baks, B. Hamel-Bissell, C. Kocot, L. Graham, R. Johnson, G. Landry, E. Shaw, A. MacInnes, and J. Tatum, "A 56.1Gb/s NRZ Modulated 850nm VCSEL-Based Optical Link," in *Optical Fiber Communication Conference/National Fiber Optic Engineers Conference 2013* (OSA, 2013), p. OW1B.5.
4. D. M. Kuchta, A. V. Rylyakov, F. E. Doany, C. L. Schow, J. E. Proesel, C. W. Baks, P. Westbergh, J. S. Gustavsson, and A. Larsson, "A 71-Gb/s NRZ Modulated 850-nm VCSEL-Based Optical Link," IEEE Photonics Technol. Lett. **27**(6), 577–580 (2015).
5. Q. Cheng, S. Rumley, M. Bahadori, and K. Bergman, "Photonic switching in high performance datacenters [Invited]," Opt. Express **26**(12), 16022–16043 (2018).
6. S. Rumley, M. Bahadori, R. Polster, S. D. Hammond, D. M. Calhoun, K. Wen, A. Rodrigues, and K. Bergman, "Optical interconnects for extreme scale computing systems," Parallel Comput. **64**, 65–80 (2017).
7. S. Rumley, D. Nikolova, R. Hendry, Q. Li, D. Calhoun, and K. Bergman, "Silicon Photonics for Exascale Systems," J. Light. Technol. **33**(3), 547–562 (2015).
8. Y. Shen, X. Meng, Q. Cheng, S. Rumley, N. Abrams, A. Gazman, E. Manzhosov, M. Glick, and K. Bergman, "Silicon Photonics for Extreme Scale Systems," J. Light. Technol. 1–1 (2019).
9. A. E. Lim, J. Song, Q. Fang, C. Li, X. Tu, N. Duan, K. K. Chen, R. P. Tern, and T. Liow, "Review of Silicon Photonics Foundry Efforts," IEEE J. Sel. Top. Quantum Electron. **20**(4), 405–416 (2014).
10. W. J. Westerveld, S. M. Leinders, K. W. A. van Dongen, H. P. Urbach, and M. Yousefi, "Extension of Marcatili's Analytical Approach for Rectangular Silicon Optical Waveguides," J. Light. Technol. **30**(14), 2388–2401 (2012).
11. W. Bogaerts and L. Chrostowski, "Silicon Photonics Circuit Design: Methods, Tools and Challenges," Laser Photonics Rev. **12**(4), 1700237 (2018).
12. K. K. Lee, D. R. Lim, L. C. Kimerling, J. Shin, and F. Cerrina, "Fabrication of ultralow-loss Si/SiO2 waveguides by roughness reduction," Opt. Lett. **26**(23), 1888–1890 (2001).
13. Z. Lu, H. Yun, Y. Wang, Z. Chen, F. Zhang, N. A. F. Jaeger, and L. Chrostowski, "Broadband silicon photonic directional coupler using asymmetric-waveguide based phase control," Opt. Express **23**(3), 3795–3808 (2015).
14. Y. Zhang, S. Yang, A. E.-J. Lim, G.-Q. Lo, C. Galland, T. Baehr-Jones, and M. Hochberg, "A compact and low loss Y-junction for submicron silicon waveguide," Opt. Express **21**(1), 1310–1316 (2013).
15. D. J. Thomson, Y. Hu, G. T. Reed, and J.-M. Fedeli, "Low Loss MMI Couplers for High Performance MZI Modulators," IEEE Photonics Technol. Lett. **22**(20), 1485–1487 (2010).
16. M. Bahadori, M. Nikdast, S. Rumley, L. Y. Dai, N. Janosik, T. V. Vaerenbergh, A. Gazman, Q. Cheng, R. Polster, and K. Bergman, "Design Space Exploration of Microring Resonators in Silicon Photonic Interconnects: Impact of the Ring Curvature," J. Light. Technol. **36**(13), 2767–2782 (2018).
17. A. H. K. Park, H. Shoman, M. Ma, S. Shekhar, and L. Chrostowski, "Ring resonator based polarization diversity WDM receiver," Opt. Express **27**(5), 6147–6157 (2019).
18. R. Ding, Y. Liu, Q. Li, Y. Yang, Y. Ma, K. Padmaraju, A. E.-J. Lim, G.-Q. Lo, K. Bergman, T. Baehr-Jones, and M. Hochberg, "Design and characterization of a 30-GHz bandwidth low-power silicon traveling-wave modulator," Opt. Commun. **321**, 124–133 (2014).

19. K. Ogawa, K. Goi, Y. T. Tan, T.-Y. Liow, X. Tu, Q. Fang, G.-Q. Lo, and D.-L. Kwong, "Silicon Mach-Zehnder modulator of extinction ratio beyond 10 dB at 10.0-12.5 Gbps," Opt. Express **19**(26), B26–B31 (2011).
20. X. Xiao, X. Li, H. Xu, Y. Hu, K. Xiong, Z. Li, T. Chu, J. Yu, and Y. Yu, "44-Gb/s Silicon Microring Modulators Based on Zigzag PN Junctions," IEEE Photonics Technol. Lett. **24**(19), 1712–1714 (2012).
21. Q. Xu, S. Manipatruni, B. Schmidt, J. Shakya, and M. Lipson, "12.5 Gbit/s carrier-injection-based silicon micro-ring silicon modulators," Opt. Express **15**(2), 430–436 (2007).
22. M. T. Wade, F. Pavanello, R. Kumar, C. M. Gentry, A. Atabaki, R. Ram, V. Stojanovic, and M. A. Popovic, "75% efficient wide bandwidth grating couplers in a 45 nm microelectronics CMOS process," in *2015 IEEE Optical Interconnects Conference (OI)* (IEEE, 2015), pp. 46–47.
23. Y. Wang, H. Yun, Z. Lu, R. Bojko, W. Shi, X. Wang, J. Flueckiger, F. Zhang, M. Caverley, N. A. F. Jaeger, and L. Chrostowski, "Apodized Focusing Fully Etched Subwavelength Grating Couplers," IEEE Photonics J. **7**(3), 1–10 (2015).
24. D. Vermeulen and C. V. Poulton, "Optical Interfaces for Silicon Photonic Circuits," Proc. IEEE **106**(12), 2270–2280 (2018).
25. A. Rao and S. Fathpour, "Second-Harmonic Generation in Integrated Photonics on Silicon," Phys. Status Solidi A **215**(4), 1700684 (2018).
26. A. H. K. Park, A. S. Ramani, L. Chrostowski, and S. Shekhar, "Comparison of DAC-less PAM4 modulation in segmented ring resonator and dual cascaded ring resonator," in *2017 IEEE Optical Interconnects Conference (OI)* (IEEE, 2017), pp. 7–8.
27. S. Palermo, K. Yu, A. Roshan-Zamir, B. Wang, C. Li, M. A. Seyedi, M. Fiorentino, and R. Beausoleil, "PAM4 silicon photonic microring resonator-based transceiver circuits," in H. Schröder and R. T. Chen, eds. (2017), p. 101090F.
28. A. Boes, B. Corcoran, L. Chang, J. Bowers, and A. Mitchell, "Status and Potential of Lithium Niobate on Insulator (LNOI) for Photonic Integrated Circuits," Laser Photonics Rev. **12**(4), 1700256 (2018).
29. A. Rao and S. Fathpour, "Heterogeneous Thin-Film Lithium Niobate Integrated Photonics for Electrooptics and Nonlinear Optics," IEEE J. Sel. Top. Quantum Electron. **24**(6), 1–12 (2018).
30. L. Chen, Q. Xu, M. G. Wood, and R. M. Reano, "Hybrid silicon and lithium niobate electro-optical ring modulator," Optica **1**(2), 112–118 (2014).
31. P. O. Weigel, J. Zhao, K. Fang, H. Al-Rubaye, D. Trotter, D. Hood, J. Mudrick, C. Dallo, A. T. Pomerene, A. L. Starbuck, C. T. DeRose, A. L. Lentine, G. Rebeiz, and S. Mookherjea, "Hybrid Silicon Photonic-Lithium Niobate Electro-Optic Mach-Zehnder Modulator Beyond 100 GHz Bandwidth," ArXiv180310365 Phys. (2018).
32. M. He, M. Xu, Y. Ren, J. Jian, Z. Ruan, Y. Xu, S. Gao, S. Sun, X. Wen, L. Zhou, L. Liu, C. Guo, H. Chen, S. Yu, L. Liu, and X. Cai, "High-Performance Hybrid Silicon and Lithium Niobate Mach-Zehnder Modulators for 100 Gbit/s and Beyond," ArXiv180710362 Phys. (2018).
33. A. Rao, A. Patil, P. Rabiei, A. Honardoost, R. DeSalvo, A. Paolella, and S. Fathpour, "High-performance and linear thin-film lithium niobate Mach–Zehnder modulators on silicon up to 50 GHz," Opt. Lett. **41**(24), 5700–5703 (2016).
34. P. Rabiei, J. Ma, S. Khan, J. Chiles, and S. Fathpour, "Heterogeneous lithium niobate photonics on silicon substrates," Opt. Express **21**(21), 25573–25581 (2013).
35. M. Zhang, C. Wang, X. Chen, M. Bertrand, A. Shams-Ansari, S. Chandrasekhar, P. Winzer, and M. Lončar, "Ultra-High Bandwidth Integrated Lithium Niobate Modulators with Record-Low Vpi," in *2018 Optical Fiber Communications Conference and Exposition (OFC)* (2018), pp. 1–3.
36. I. Krasnokutska, J.-L. J. Tambasco, and A. Peruzzo, "Large free spectral range microring resonators in lithium niobate on insulator," 6 (n.d.).
37. C. Wang, M. Zhang, B. Stern, M. Lipson, and M. Lončar, "Nanophotonic lithium niobate electro-optic modulators," Opt. Express **26**(2), 1547–1555 (2018).
38. I. Krasnokutska, J.-L. J. Tambasco, X. Li, and A. Peruzzo, "Ultra-low loss photonic circuits in lithium niobate on insulator," Opt. Express **26**(2), 897–904 (2018).
39. I. of E. Engineers, *Properties of Lithium Niobate* (IET, 2002).
40. L. Cai, A. Mahmoud, and G. Piazza, "Low-loss waveguides on Y-cut thin film lithium niobate: towards acousto-optic applications," Opt. Express **27**(7), 9794 (2019).
41. M. Mahmoud, S. Ghosh, and G. Piazza, "Lithium Niobate on Insulator (LNOI) Grating Couplers," in *CLEO: 2015 (2015), Paper SW4I.7* (Optical Society of America, 2015), p. SW4I.7.

42. M. Mahmoud, L. Cai, C. Bottenfield, and G. Piazza, "Lithium Niobate Electro-Optic Racetrack Modulator Etched in Y-Cut LNOI Platform," IEEE Photonics J. **10**(1), 1–10 (2018).
43. J. Jian, P. Xu, H. Chen, M. He, Z. Wu, L. Zhou, L. Liu, C. Yang, and S. Yu, "High-efficiency hybrid amorphous silicon grating couplers for sub-micron-sized lithium niobate waveguides," Opt. Express **26**(23), 29651–29658 (2018).
44. M. Bahadori, A. Kar, Y. Yang, A. Lavasani, L. Goddard, and S. Gong, "High-Performance Integrated Photonics in Thin Film Lithium Niobate Platform," in *Conference on Lasers and Electro-Optics* (2019).
45. A. Kar, M. Bahadori, S. Gong, and L. Goddard, "Realization of alignment-tolerant grating couplers for z-cut thin-film lithium niobate [under review]," Opt. Express (2019).
46. M. Bahadori, M. Nikdast, Q. Cheng, and K. Bergman, "Universal Design of Waveguide Bends in Silicon-on-Insulator Photonics Platform," J. Light. Technol. 1–1 (2019).
47. Z. Hu and Y. Y. Lu, "Computing Optimal Waveguide Bends With Constant Width," J. Light. Technol. **25**(10), 3161–3167 (2007).
48. H. Shen, M. Hkan, S. Xiao, and M. Qi, "Reducing mode-transition loss in silicon-on-insulator strip waveguide bends," in *2008 Conference on Lasers and Electro-Optics* (IEEE, 2008), pp. 1–2.
49. O. Reshef, M. G. Moebius, and E. Mazur, "Extracting losses from asymmetric resonances in micro-ring resonators," J. Opt. **19**(6), 065804 (2017).
50. W. Bogaerts, P. De Heyn, T. Van Vaerenbergh, K. De Vos, S. Kumar Selvaraja, T. Claes, P. Dumon, P. Bienstman, D. Van Thourhout, and R. Baets, "Silicon microring resonators," Laser Photonics Rev. **6**(1), 47–73 (2012).